**Current knowledge and future research opportunities for modeling annual crop mixtures. A review.**

Noémie Gaudio* 1; Abraham J. Escobar-Gutiérrez 2; Pierre Casadebaig 1; Jochem B. Evers 3; Frédéric Gérard 4; Gaëtan Louarn 2; Nathalie Colbach 5; Sebastian Munz 6; Marie Launay 7; Hélène Marrou 8; Romain Barillot 2; Philippe Hinsinger 4; Jacques-Eric Bergez 1; Didier Combes 2; Jean-Louis Durand 2; Ela Frak 2; Loïc Pagès 9; Christophe Pradal 10,11; Sébastien Saint-Jean 12; Wopke Van Der Werf 3; Eric Justes 8

1 AGIR, University of Toulouse, INRA, Castanet-Tolosan, France
* noemie.gaudio@inra.fr
2 INRA, URP3F, Equipe Ecophysiologie des Plantes Fourragères, Le Chêne – RD 150, BP 6, F-86600 Lusignan, France
3 Centre for Crop Systems Analysis, Wageningen University and Research, Wageningen, the Netherlands
4 Eco&Sols, Univ. Montpellier, CIRAD, INRA, IRD, SupAgro, Place Viala, 34060, Montpellier, France
5 Agroécologie, AgroSup Dijon, INRA, Univ. Bourgogne Franche-Comté, F-21000 Dijon, France
6 Department of Agronomy, Institute of Crop Science, University of Hohenheim, Stuttgart, Germany
7 INRA, US1116 AgroClim, Agroparc 84914 Avignon cedex 9, France
8 CIRAD, UMR SYSTEM, Univ. Montpellier, CIHEAM-IAMM, CIRAD, INRA, Montpellier Supagro, 34398 Montpellier, France
9 INRA, Centre PACA, UR PSH 1115, Agroparc 84914 Avignon cedex 9
10 AGAP, Univ. Montpellier, CIRAD, INRA, SupAgro, Montpellier, France
11 CIRAD, AGAP, and INRIA Zenith, Univ Montpellier, Montpellier, France
12 UMR ECOSYS INRA, AgroParisTech, Université Paris-Saclay, 78850 Thiverval-Grignon, France

**Abstract**

Growing mixtures of annual arable crop species or genotypes is a promising way to improve crop production without increasing agricultural inputs. To design optimal crop mixtures, choices of species, genotypes, sowing proportion, plant arrangement, and sowing date need to be made but field experiments alone are not sufficient to explore such a large range of factors. Crop modeling allows to study, understand and ultimately design cropping systems and is an established method for sole crops. Recently, modeling started to be applied to annual crop mixtures as well.

Here, we review to what extent crop simulation models and individual-based models are suitable to capture and predict the specificities of annual crop mixtures. We argued that: 1) The crop mixture spatio-temporal heterogeneity (influencing the occurrence of ecological processes) determines the choice of the modeling approach (plant or crop centered). 2) Only few crop models (adapted from sole crop models) and individual-based models currently exist to simulate annual crop mixtures. 3) Crop models are mainly used to address issues related to crop mixtures management and to the integration of crop mixtures into larger scales such as the rotation, whereas individual-based models are mainly used to identify plant traits involved in crop mixture performance and to quantify the relative contribution of the different ecological processes (niche complementarity, facilitation, competition, plasticity) to crop mixture functioning.

This review highlights that modeling of annual crop mixtures is in its infancy and gives to model users some important keys to choose the model based on the questions they want to answer, with awareness of the strengths and weaknesses of each of the modeling approaches.

**Content**

# 1. Introduction

The discipline of crop modeling emerged in the 1950s (Keating and Thorburn 2018) and is recognized as a crucial and operational tool to support improvement of cropping systems. One of its main strengths is that it enables multiple combinations of environments (climate and soil), genotypes, and agricultural practices to be explored and compared based on current knowledge of crop functioning in interaction with the environment (Bergez et al. 2010; Boote et al. 2013; Jeuffroy et al. 2014). A large scientific community supports crop modeling in international consortia, such as AgMIP ("Agricultural Model Intercomparison and Improvement Project", Rosenzweig et al. 2013), to develop, improve, and evaluate models. For mainstream sole cropping systems (i.e. one genotype), crop modeling has been used to identify factors that limit crop productivity (e.g. Brisson et al. 2010), assess environmental impacts of crops (e.g. Liu et al. 2016), decrease inputs through improved management decisions (e.g. maize crop irrigation, Bergez et al. 2001; wheat fertilization, Chatelin et al. 2005), and aid breeding of new cultivars (reviewed by Chenu et al. 2017). At the field scale, sole crops are relatively homogeneous, especially when external inputs are used. Currently, crop modeling is facing challenges because the agroecological transition involves changes in agricultural practices (higher plant diversity, lower inputs, less tillage, biological regulation, tolerating residual weeds, etc.), that result in agroecosystems with greater complexity. Increasing public concerns about human and environmental health issues related to intensive agricultural systems over the past 70 years in Western industrialized countries have prompted policy makers and scientists to search for alternative management strategies, such as more diversified cropping systems (Duru et al. 2015).

Increasing plant diversity in agriculture is suggested as a pathway towards more resilient and sustainable production systems (Lin 2011; Altieri et al. 2015; Raseduzzaman and Jensen 2017). At the farm scale, diversification can occur by diversifying crops in rotations (Gliessman 2014; Reckling 2016) and integrating agroecological pest (i.e. any organism harmful for crops) management to decrease pesticide use at farm and landscape scales (Lechenet et al. 2017; Hatt et al. 2018). At the field scale, plant diversity can increase using within-field mixtures of at least two annual crop species (i.e. intercropping, Figure 1, Vandermeer 1989) or genotypes of the same species, both hereafter referred to as "annual crop mixtures". Studies have demonstrated advantages of annual crop mixtures compared to their corresponding sole crops (reviewed by Lithourgidis et al. 2011; Bedoussac et al. 2015; Brooker et al. 2015; Li-li et al. 2015 for species diversity and Zeller et al. 2012; Creissen et al. 2016; Barot et al. 2017 for genotype diversity). Recent reviews and meta-analyses summarized the main ecosystem services they can deliver (Kiaer et al. 2009; Malézieux et al. 2009; Kremen and Miles 2012; Ehrmann and Ritz 2014; Altieri et al. 2015; Barot et al. 2017; Duchene et al. 2017; Raseduzzaman and Jensen 2017), highlighting their benefits for crop production (e.g. yield quality, quantity and stability), improvement of soil biogeochemistry (e.g. fertility, water flow regulation), improvement of biological pest control, and climate regulation by mitigating greenhouse gas emissions.

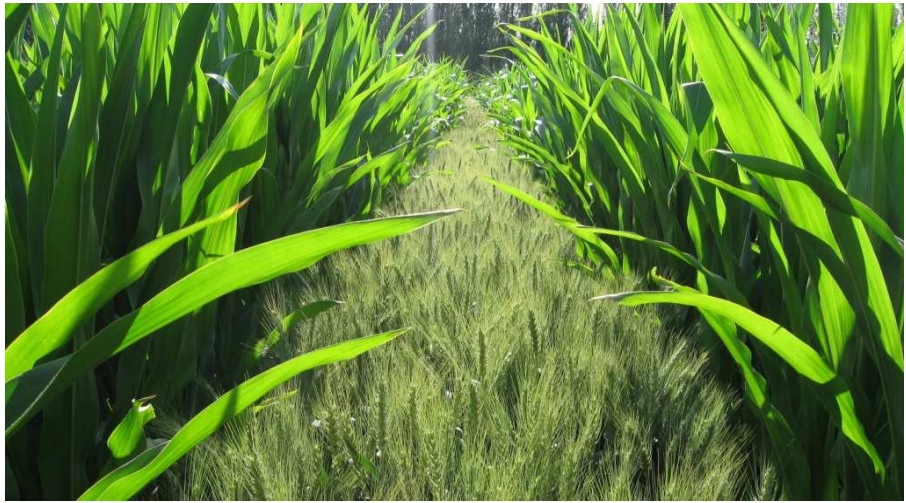

Figure 1. Relay-strip intercrop of maize and wheat in the Netherlands (Source: J. Evers, Wageningen University and Research).

Despite the potential of annual crop mixtures, they are under-represented in modern agriculture (Machado 2009; Costanzo and Bàrberi 2014). Like for sole crops, modeling could provide management guidelines to promote and optimize their use. More specifically, designing productive and resilient crop mixtures is challenging, particularly with respect to i) the choice and proportions of species and genotypes (Baxevanos et al. 2017) and ii) their spatio-temporal arrangement in the field (Wezel et al. 2014). Both are intended to optimize plant-plant complementarities and positive interactions while minimizing competition to best respond to the existing environment in space and time (Hinsinger et al. 2011; Brooker et al. 2015, 2016). Nevertheless, if crop mixtures are not adequately managed, they may not provide more benefits than sole crops (Brooker et al. 2015). In response, models can analyze and promote the use of crop mixtures (Malézieux et al. 2009) by guiding the design of plant arrangement, the duration of temporal overlap, and plant characteristics, as well as predicting crop mixtures behavior and the services they deliver, and increasing understanding of their functioning. Additionally, breeding, currently used to increase sole crop performance, may not be optimal for crop mixtures, especially due to trait plasticity (Zhu et al. 2016; Kiaer and Boesen 2018). Therefore, models could also help identify ideotypes for crop mixtures and quantify the relative importance of trait values and plasticity (Litrico and Violle 2015).

Modeling of annual crop mixtures is under development, and the few models that currently exist focus mainly on abiotic resource partitioning. Two main modeling approaches are currently used: i) process-based crop models, in which crop characteristics are represented at the field scale (Launay et al. 2009; Knörzer et al. 2011; Fayaud et al. 2014; Munz et al. 2014; Gou et al. 2017b), and ii) individual-based models (IBMs), in which each plant is represented individually at various levels of architectural realism (Garcia-Barrios et al. 2001; Potting et al. 2005; Postma and Lynch 2012; Barillot et al. 2014b; Colbach et al. 2014b; Zhu et al. 2015). These models have difficulty representing the specific characteristics and complexity of crop mixtures because they may not completely capture certain processes (complementarity, facilitation, resource partitioning, etc.) that influence crop mixture performance. Therefore, it is necessary to assess whether existing models can address the issues of annual crop mixtures. If not, existing models should be adapted or new models and modeling approaches should be developed (Affholder et al. 2012) to represent concepts of crop mixtures that differ fundamentally from those of sole crops.

This paper aims at assessing suitable modeling approaches to the issues of annual crop mixtures and at highlighting issues that cannot be addressed by current models. Annual crop mixtures are classified based on spatio-temporal heterogeneity and the assumption that agroecosystem heterogeneity influences the types of models and their corresponding formalisms. A universal model applicable to all agroecosystems is not feasible; therefore, model development depends on the system and the issues considered (Sinclair and Seligman 1996; Affholder et al. 2012). The key ecological processes within crop mixtures are discussed, focusing on the spatial scale at which they occur and their relative importance as a function of the type of crop mixture. Then, the paper focuses on the ability of existing models (crop models and IBMs) to represent the types of crop mixtures identified. The main features and applications of each modeling approach are discussed. Our review focuses particularly on analyzing strengths, weaknesses, and the complementarity of existing models as an initial step to help improving them in the future.

## 2. Classification of annual crop mixtures focusing on the spatial scale of ecological processes

We focused on mixtures of at least two annual cash crops in which the spatio-temporal arrangement is particularly important for both crops, thus excluding mixtures of a cash crop (usually with a defined spatial arrangement) and a broadcast undersown cover crop (e.g. Arim et al. 2006).

### 2.1. Spatio-temporal heterogeneity defines the types of annual crop mixtures

Spatio-temporal arrangements and plant components (species/genotypes) of crop mixtures are diverse (Gaba et al. 2015), resulting in varying levels of heterogeneity within the crop. Thus, a continuum exists within annual crops (Malézieux et al. 2009), ranging from high crop homogeneity − sole crops of a single genotype or genotype mixtures where genotypes differ only in resistance to pests (Finckh et al. 2000; Tooker and Frank 2012) − to high heterogeneity − crop mixtures of species or genotypes with contrasting morphology and phenology (Essah and Stoskopf 2002; Barillot et al. 2014a; Wang et al. 2015; Montazeaud et al. 2017; Vidal et al. 2018). Morphology is described by plant architecture (Godin 2000; Bucksch et al. 2017), while phenology is related to the key stages of plant development (from emergence to maturity, Lieth 1974), keeping in mind that genotypes and species can still differ by inherent physiological differences not related to morphology and phenology (e.g. secondary metabolites deterring attacks from herbivores).

Three types of annual crop mixtures were defined according to their level of heterogeneity (Figure 2): type A, in which plant components have similar phenologies and morphologies; type B, in which plant components have contrasting morphologies but relatively similar phenologies; and type C, in which plant components have contrasting phenologies and morphologies.

Within this classification, spatio-temporal arrangements of crop mixtures meet morphological and phenological heterogeneity requirements for harvest. For instance, relay-strip intercrops of wheat and maize (type C) combine two species with highly contrasting phenologies; therefore, they are usually sown on different dates in alternate strips to maximize resource use, which results in temporally separated harvests (Li et al. 2001; Gao et al. 2010; Wang et al. 2015; Gou et al. 2017a). It is also the case with some genotypes mixtures despite these agroecosystems are rarely used in agriculture, e.g. relay-strip mixture of spring- and summer-maize (Ning et al. 2012). In contrast, homogeneous genotypes of rice (type A, Han et al. 2016), morphologically different genotypes of wheat (type B, Vidal et al. 2018) or barley and pea (type B, Hauggaard-Nielsen et al. 2009), have phenologies similar enough to be harvested at the same time. Thus, the sowing arrangements most often used are within-row and alternate-row mixtures, which create close plant-plant interactions.

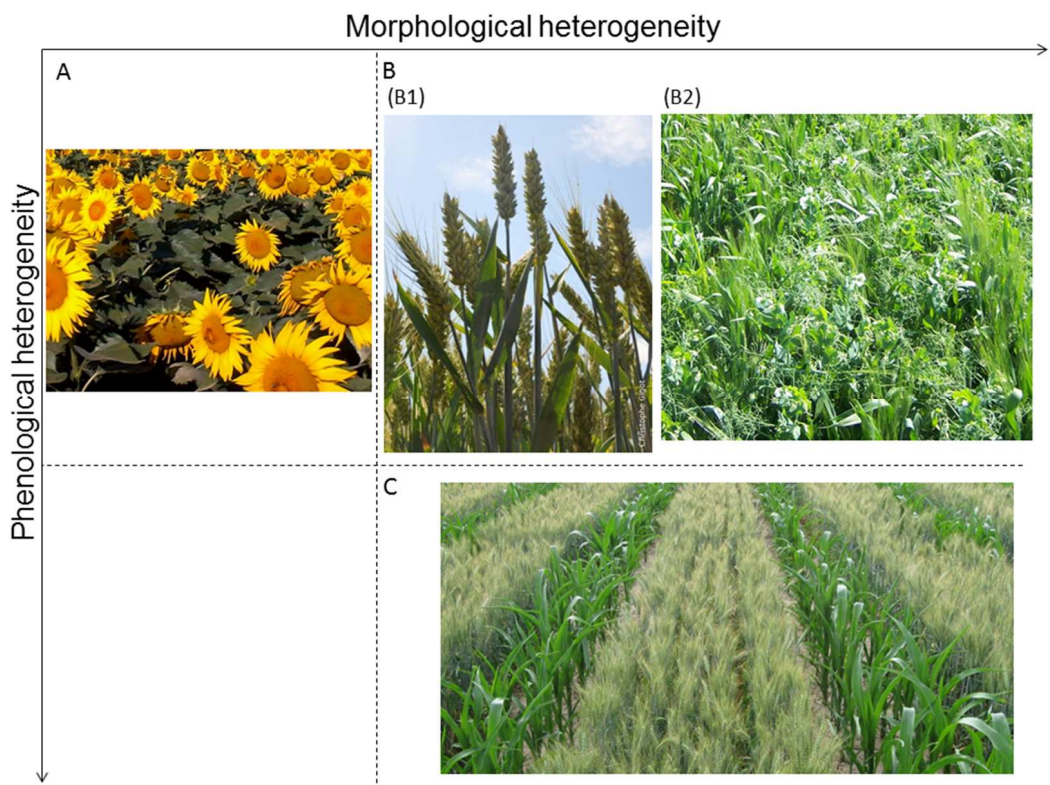

Figure 2. Classification of annual crop mixtures based on a double gradient of phenological and morphological heterogeneity. Morphological heterogeneity is related to plant architecture, while phenological heterogeneity is related to the level of asynchrony of key developmental stages. Photos: (A) within-row mixture of three genotypes of sunflower with contrasting degrees of pest resistance, France (Source: C. Bonnet, INRA AGIR); (B1) within-row mixture of two genotypes of wheat, France (Source: C. Gigot, INRA AgroParisTech EcoSys); (B2) alternate-

row mixture of wheat and pea, France (Source: L. Bedoussac, ENSFEA, INRA AGIR); (C) relay-strip intercrop of wheat and maize, the Netherlands (Source: J. Evers, Wageningen University and Research).

2.2. Relative importance of ecological processes in the three types of annual crop mixtures

The ability of crop mixtures to use abiotic resources depends in part on several ecological processes (reviewed by Brooker et al. 2015; Barot et al. 2017) which often overlap throughout the growing cycle. Niche complementarities (Macarthur and Levins 1967) occur through niche differentiation over time (Yu et al. 2015; Zhang et al. 2017; Dong et al. 2018) and/or space (Zhang et al. 2014; Montazeaud et al. 2017), and by using different forms of the same resource, as illustrated for nitrogen (N) uptake by cereal-legume mixtures (e.g. Hauggaard-Nielsen et al. 2009) or for the preferential use of nitrate or ammonium (Boudsocq et al. 2012); phenotypic plasticity (Nicotra et al. 2010) which can contribute greatly to niche complementarity (Callaway et al. 2003; Ashton et al. 2010; Zhu et al. 2015), e.g. root traits are extremely plastic (Pagès 2011) and can behave differently when growing in mixtures (e.g. Gonkhamdee et al. 2010 for maize intercropped with rubber tree) so that if models exclude plastic responses in root development as an emergent property, simulating the distribution of roots of crop mixtures based on that of single crops could be irrelevant; positive interactions (i.e. facilitation, Callaway 1995), e.g. one species can increase phosphorous (P) availability in its rhizosphere and ultimately in the rhizosphere of a neighboring species, provided that the two root systems are intermingled (Gunes et al. 2007; Betencourt et al. 2012; Li et al. 2014; Bargaz et al. 2017). In addition, the selection effect, also named the sampling effect (Loreau and Hector 2001; Barot et al. 2017), could also lead to overyielding. This is due to the fact that some genotypes or species grow particularly well in given cropping conditions. Thus, simulation models taken into account plant phenology and autecology would be able to consider and quantify this effect.

These ecological processes occur at different spatial scales and with different intensities depending on cropping conditions. In a given environment, their relative importance varies according to the type of annual crop mixture, because spatio-temporal arrangement determines the immediate neighborhood of a target plant (Stoll and Weiner 2000, Table 1). Regardless of the crop mixture type, and even for sole crops, phenotypic plasticity is crucial for adapting to the local environment and its constraints. The three types of crop mixtures can be considered a logical sequence (type A → type B → type C) by successively adding processes and complexity to account for.

In mixtures with homogeneous genotypes (type A), all individual plants experience approximately the same local abiotic environment, ignoring i) spatial heterogeneity caused by emergence conditions and pests; and ii) border effects, such as those of surrounding hedges or grass strips. For type A crop mixtures, above- and belowground competition may be nearly the same for all plants and depend on the density of a given crop in a given environment. In general, no facilitation or niche complementarity for abiotic resources can occur. However, facilitation-like mechanisms can occur through a decrease in pests due to contrasting degrees of resistance (Barot et al. 2017). Predicting this specific process and its dynamics would require distinguishing each plant.

Next in the sequence, in alternate-row and within-row mixtures (type B), crops with contrasting morphology are closely mixed. Therefore, above- and belowground interactions (i.e. competition and facilitation) and complementarity for resource use are important for predicting crop mixture functioning and performances, such as yield and other ecosystem services. For example, modeling belowground niche complementarity in space, Postma and Lynch (2012) predicted that in closely spaced mixtures of maize-bean-squash (type B), different root architectures (Figure 3) allow crop mixtures to forage for nutrients throughout the soil profile more efficiently than a single genotype/species. The functioning of a target plant is determined by its neighborhood, with a high probability of being close to a neighbor from another species or genotype according to the relative density of the plant components. Alternate-row mixtures with contrasting degrees of resistance strengthen facilitation against pests, since non-host crop rows serve as physical barriers to pest movement between host rows (Finckh et al. 2000). The scale (individual or population) at which these ecological processes should be considered, however, depends on the specific issue addressed.

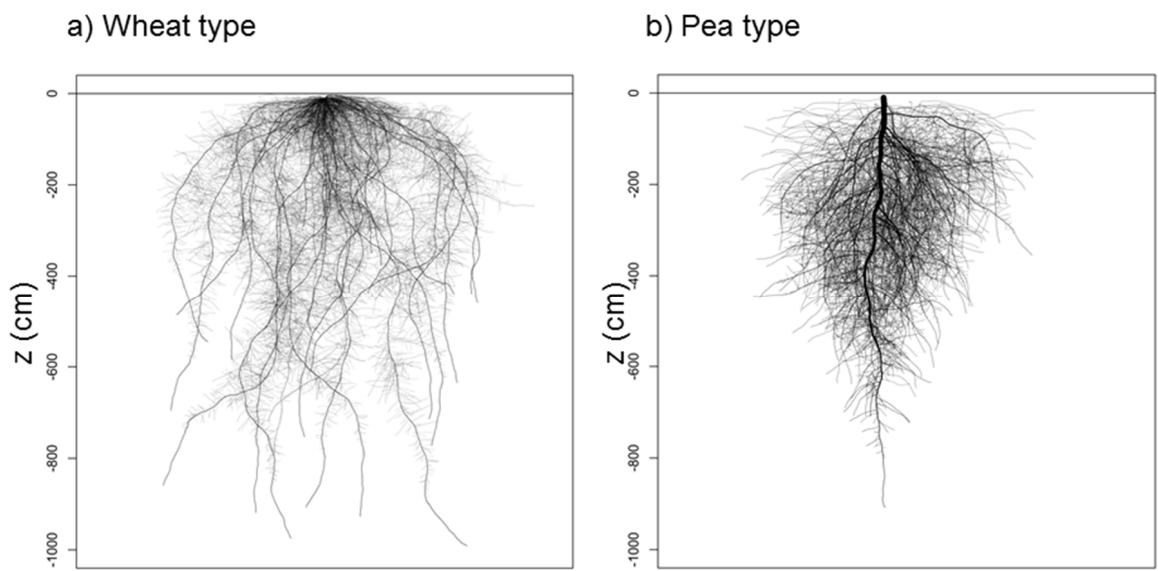

Figure 3. Illustration of two root systems with contrasting architecture: (a) wheat, with a branching root system, and (b) pea, with a taproot system, simulated with the root model ArchiSimple (Pagès et al. 2014).

The main additional feature that distinguishes relay-strip crop mixtures (type C) from the other two types is the strong border-row effect between two strips, due to interactions between the two crops as they grow. Competition for light (Yang et al. 2017) or belowground resources (Lv et al. 2014) can predominate, depending on the plant components, their water and nutrient status, and their spatio-temporal arrangement (e.g. number of rows, orientation). However, it is assumed that belowground facilitation and complementarity for N, that occur at local scales, can have a strong influence only at the border between two strips, i.e. the wider the strip, the lower the effects. For belowground resources, especially P, the strongest effect of increased P use in faba bean-maize relay-strip intercrops was linked to decomposition of roots of the species harvested first (Li et al. 2003) and to slight direct facilitation during co-growth. As Raynaud et al. (2008) modeled, roots must be intimately mixed for P facilitation given the small scale of the underlying rhizosphere processes (Hinsinger et al. 2011). Temporal niche complementarity is the main ecological process involved in relay-strip crop mixtures (Yu et al. 2015). Several studies showed that wheat overyielding in wheat-maize relay-strip intercrops was due mainly to increased growth of border-row plants, while inner-row plants behaved like those in wheat sole crops (Knörzer et al. 2010; Gou et al. 2016; Zhu et al. 2016). Wheat border rows received more light than inner rows before maize emerged, resulting in greater light use efficiency in crop mixtures than in sole crops. This highlights the need to consider the spatial pattern of relay-strip crop mixtures explicitly. Whatever the crop mixture type, the local environment determines which functional complementarities between plants sustain growth and production under resource limitation. For instance, when N, water or P is the main limiting resource, then complementarity for N use between legume and non-legume species (Hauggaard-Nielsen et al. 2009), for water use between C3 and C4 species (Mao et al. 2012), or for P use between P- and non-P-mobilizing species (Li et al. 2014) becomes the most advantageous, respectively. This first environmental filter (Keddy 1992) guides the choice of the mixture components and occurs at the species level and not the genotype level because it is influenced by general properties of a species that are shared by all of its genotypes, thus defining coarse plant functional types (Ustin and Gamon 2010).

| | | Type A | Type B | Type C |
|---|---|---|---|---|
| Above- and belowground competition (intra- and interspecific / genotypes) | occurrence | **x** | **x** | **x** |
| | influenced by | Sowing density. | Sowing density, proportion of each mixture component, and competitive ability (Goldberg 1990). | Sowing density, proportion of each mixture component, competitive ability, and row position in a given strip. |
| Phenotypic plasticity | occurrence | **x** | **x** | **x** |
| | influenced by | Each mixture component sensitivity to the environment, their response range, and direct neighborhood. | | |
| Spatial complementarity for belowground resources | occurrence | | **x** | **x** |
| | influenced by | | Root system architecture of each mixture component, and direct neighborhood. | |
| Complementarity for resource use | occurrence | | **x** | **x** |
| | influenced by | | Functional types characterizing each mixture component. | |
| Facilitation | occurrence | | **x** | **x** |
| | influenced by | | Functional types characterizing each mixture component, and the direct neighborhood. | |
| Representation from a target plant point of view (direct neighborhood) | | A given plant is surrounded by plants with similar characteristics. All plants encounter almost the same environment and neighborhood, *i.e.* one average plant can be considered as representative of the whole crop. This is true except for plants located at the border of the field. | Each plant 1 is at least surrounded by one plant 2 in the case of alternate row mixtures and with a high probability in the case of within-row mixtures. This probability depends on the relative proportion of the two plant types. | The direct neighborhood of a target plant 1 growing in a given row can be either plant 1 or plant 2 depending on the position of the row within a given strip. Plants growing on the border rows experience a reduced neighborhood during part of their growing cycle. |
| Schematic representation | | 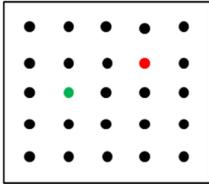 | Alternate rows<br>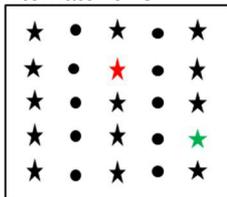<br>Within-row<br>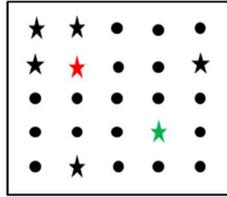 | 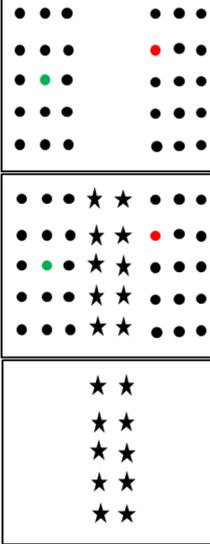 |

Table 1. Main ecological processes involved in abiotic resource partitioning in the three types of annual crop mixtures (types A, B, C) in a given environment. Their occurrence in the three types of crop mixtures and the main influencing factors are described. For the schematic representation: Crop 1 = black circle; Crop 2 = black star; Target plant 1 = green; Target plant 2 = red.

# 3. General features of current modeling approaches

## 3.1. Crop models

Crop models were developed to simulate soil-plant-atmosphere interactions by considering environmental variables (climate, soil, and management practices), species or genotype-specific traits, and their response to the environment (reviewed by Boote et al. 2013; Jeuffroy et al. 2014 for sole crops). They accurately simulate observed genotype-environment interactions for a range of genotypes in multi-environment trials (e.g. for sunflower, Casadebaig et al. 2016a). One main strength of these models is that they consider the effects of several abiotic stresses (e.g. water, N, temperature) and their interactions on crop performance (Figure 4, illustrates how a variation in plant trait values impacts crop yield, in a large diversity of cropping conditions), providing a quantitative estimate at a relevant scale (e.g. yield.ha$^{-1}$). Therefore, they are potentially relevant for addressing the performance of crop mixtures compared to that of sole crops under a variety of environmental conditions, such as drought or nutrient limitations (Hughes and Stachowicz 2004; Holmgren and Scheffer 2010; Hautier et al. 2014).

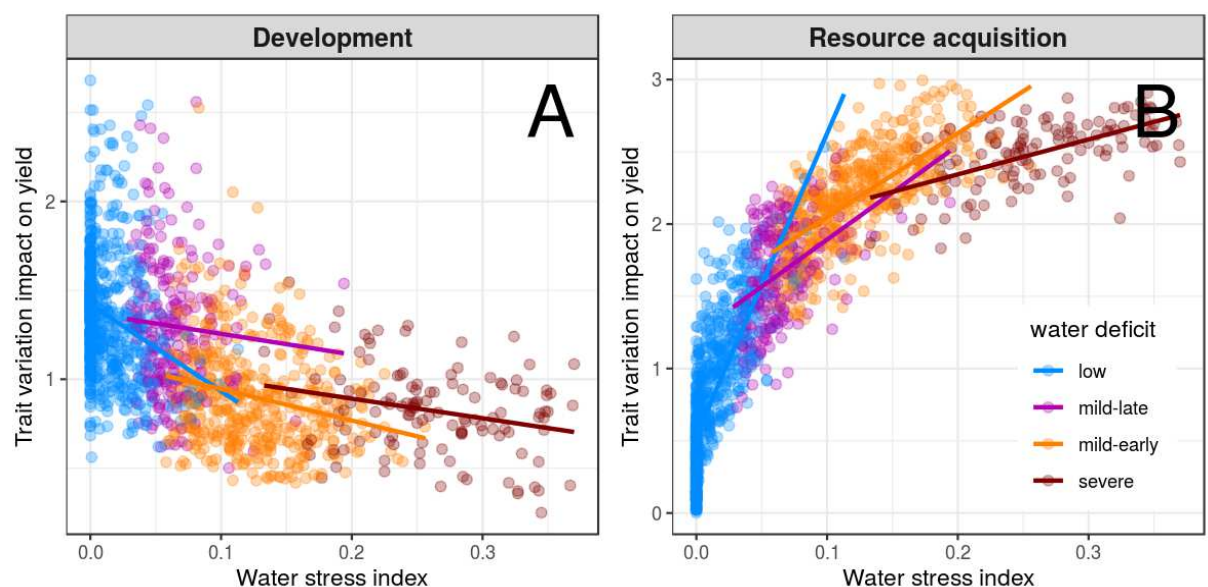

Figure 4. Sensitivity index of crop yield for selected traits involved in crop development (A) and resource acquisition (B) relative to seasonal water stress indices. Crop modeling and simulation can be used to quantify in silico the impact of a variation in plant traits on crop yield, for diverse cropping conditions. The results from a sensitivity analysis are illustrated for a subset of two traits for the wheat crop in Australia: the thermal time required to reach floral initiation (*Development*), the water extractability by roots (*Resource acquisition*). Points correspond to sampled cropping conditions (1500 combinations of sites x year x management options). Colors indicate representative drought-pattern environment types occurring in the Australian wheat belt (Chenu et al. 2013). Lines represent linear regressions fitted by environment types. Source: Casadebaig et al. (2016b).

Most crop models, however, initially simulated only sole crops, assuming a continuous canopy across the field and often ignoring spatial heterogeneity. Many models represent the entire canopy as a single leaf (in "big leaf" models) and use the Beer-Lambert law to simulate its light interception (Sinoquet et al. 1990; Monsi and Saeki 2005). Some models represent vertical spatial heterogeneity by distinguishing several canopy layers (e.g. SUNFLO, Casadebaig et al. 2011) or using leaf-to-canopy approaches for light interception and carbon assimilation (Boote et al. 2013). The underlying assumption that the canopy is homogeneous vertically and continuous horizontally is often a robust approach to simulate sole crop growth (Chenu et al. 2017). Competition is influenced by density (Evers et al. 2006; Baccar et al. 2011; Dornbusch et al. 2011), with the caveat that density dependence is fixed or weakly parameterized in many crop models (e.g. Brisson et al. 2008).

An initial strategy to simulate crop mixtures involved modifying how one crop senses its environment by altering environmental variables according to predictions of another sole crop model (i.e. successively running individual sole crop models, Monzon et al. 2007) or external calculations (Knörzer et al. 2010; Munz et al. 2014). After calculating the impact of crop 1 on the environment, the modified environmental variables are used to simulate crop 2. This methodology is used when the study focuses only on crop 2, with no retroaction on crop 1.

A few crop models, such as APSIM (Holzworth et al. 2014), CROPSYST (Singh et al. 2013), FASSET (Berntsen et al. 2004) and STICS (Brisson et al. 2004), added sub-modules to represent competition for abiotic resources by annual crop mixtures (reviewed by Chimonyo et al. 2015). They assumed that the canopy is composed of two species instead of one, either ignoring the spatial configuration of the two crops' canopies (i.e. by simulating a shared canopy; APSIM, CROPSYST, FASSET) or explicitly representing it. The latter approach, restricted to situations in which the two crops were side by side (e.g. in alternate rows), was used only in a few models, e.g. STICS. These models used a "hedgerow" approach: a 2D representation of row geometry to calculate light interception by row crop canopies. Simulation of an alternate-row crop mixture of leek and celery using the two approaches (homogeneous horizontal leaf area distribution vs. row geometry) demonstrated, however, that they could simulate light partitioning equally well (INTERCOM, Baumann et al. 2002). These results indicate the need for more frequent assessment of the advantages of making models more complex. All these models seem to simulate canopy growth, particularly leaf area index, less accurately, but simulate nutrient cycling well (Berntsen et al. 2004; Corre-Hellou et al. 2007, 2009).

More recently, crop models were specifically developed to simulate annual crop mixtures in alternate strips (Wang et al. 2015; Gou et al. 2017b; Liu et al. 2017), with particular focus on the temporal sequence of their phenological phases. These models distinguish the phases during which plant components grow separately (equivalent to a sole crop, but interspersed with strips of empty space due to the space needed for the companion species) from phases of co-growth (with two plant species being present simultaneously). These models calculate the light partitioning between the plant components in the intercrop based on a mathematical model that takes into account the block structure and strip-path geometry of the crop mixture (Goudriaan 1977; Pronk et al. 2003). They give quite different results than a standard Beer-Lambert law for light interception in a mixed canopy that does not take the block structure and strip-way geometry into account (Gou et al. 2017b). They thus represent temporal niche complementarity for light use, but the border-row effect is ignored because each strip is considered as homogeneous cover, i.e. no distinction between inner and border rows of a strip.

While light competition is always represented (appropriately or not) in models of crop mixtures, competition for belowground resources taken up by roots is not, which remains a main weakness of these models. Belowground competition can have a larger effect than aboveground competition (Wilson 1988), especially when inputs decrease. Most crop models are based on the Monteith equation (1977), which considers intercepted light the driving force of crop growth. However, most crop models do not simulate root biomass or density in soil layers, and represent soil and its spatial heterogeneity coarsely. Consequently, direct interactions between rooting systems are virtually ignored in some crop models adapted for annual crop mixtures (e.g. Gou et al. 2017b), but can be accounted for indirectly. Root systems can be represented with species- or genotype-specific parameters, such as root depth and density distribution throughout the soil profile, e.g. STICS or BISWAT (Bertrand et al. 2018). Thus, soil resources taken up by one crop change the abiotic stress and thus the specific growth response of the second crop. When the soil is represented as several layers, responses can be simulated as preferential root growth in soil layers with higher water or nutrient content or as an increase in other processes, such as biological $N_2$ fixation for legumes instead of soil mineral N uptake. In this way, basic spatial niche complementarity for water and N between two plant components can be indirectly simulated.

3.2. Individual-based models

IBMs of plant communities have long been useful in ecology to test theories and hypotheses (Huston et al. 1988; Railsback and Grimm 2012; Evers et al. 2018a). They include population dynamics models, with simulations of annual (e.g. Zhu et al. 2015) or multi-annual growth cycles (and thus multiple generations of plants, e.g. Colbach et al. 2014a). IBMs represent populations made up of individuals, usually spatially explicit, that may differ from one another. Most IBMs explicitly represent local interactions, individual variability, and the heterogeneity of resource partitioning within the field and the crop (reviewed by Grimm and Railsback 2005; Berger et al. 2008; Vos et al. 2010; Dunbabin et al. 2013). Plant architecture is represented at varying levels of detail (e.g. sub-organ, organ, phytomer, axis, and plant for shoot architecture; root segment and root system for root architecture), with the smallest-scale IBMs integrating the topological and spatial organization of a plant's modular structure, i.e. functional-structural plant models (FSPMs, Fourcaud et al. 2008; Vos et al. 2010; DeJong et al. 2011). Regardless of the degree of detail used to represent the plant, IBMs i) require many parameters, which makes it difficult to generalize their use; ii) require laborious data collection; and iii) have high computational cost. Moreover, a recurring difficulty in modeling concerns the scaling up from one hierarchical level of organization to another and whether such levels are strongly identified in biological systems (Potochnik and McGill 2012). Given the complexity of most IBMs, they cannot be easily scaled up to address issues at large spatial scales (larger than the plot level) over long periods of time (Purves et al. 2008). Consequently, a great difficulty lies, for example, in scaling up from an IBM to an estimate of annual crop yield.

IBMs have been used in forestry and agronomy to predict dynamics of cultivated plant communities, such as multispecies grasslands (Soussana et al. 2012; Louarn et al. 2014; Durand et al. 2016; Faverjon et al. 2018), mixed forests (Liu and Ashton 1995; Chave 1999; Pérot and Picard 2012), and crop-weed mixtures (Colbach et al. 2014b; Evers and Bastiaans 2016; Renton and Chauhan 2017). IBMs have rarely been applied to annual crop mixtures, however, in part because so few IBMs exist that can simulate them. We identified three IBMs (Garcia-Barrios et al. 2001; Potting et al. 2005; FLORSYS, Colbach et al. 2014b) that simulate only aboveground interactions for light and four FSPMs that simulate only above- (Figure 5, Barillot et al. 2014b; Zhu et al. 2015) or belowground (SimRoot, Postma and Lynch 2012; Min3P-ArchiSimple, Gérard et al. 2018) compartments, even though FSPMs have considerable potential to simulate species mixtures (Evers et al. 2018b). Aboveground FSPMs mainly address competition for light but recent efforts have been made to include other processes, e.g. Barillot et al. (2018) used a comprehensive FSPM to assess C-N acquisition, allocation and grain yield of theoretical wheat genotype mixtures of plants that only differ in leaf inclination.

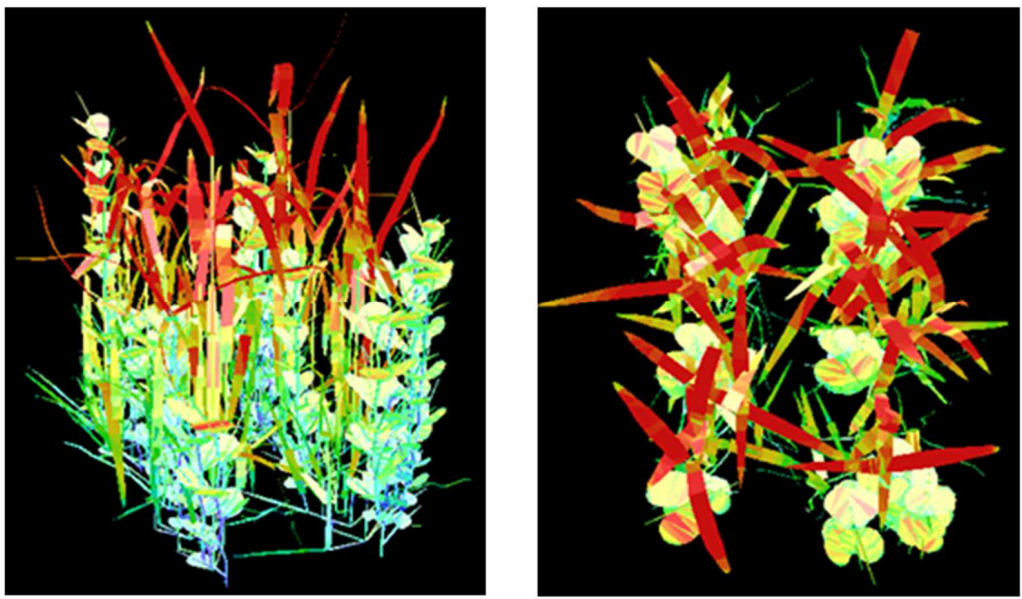

Figure 5. Illustrations of (left) horizontal and (right) vertical views of a virtual wheat-pea mixture. The color gradient (from blue to red) indicates the amount of light intercepted by plant organs. Source: Barillot (2012).

While all IBMs account to some extent for plant geometry (as a volume or surface), IBMs that are FSPMs also represent plant topology and detailed 3D architecture explicitly. Once plant geometry is calculated from input parameters (e.g. phyllotaxis, leaf and tiller angles and curvature), microclimate and phylloclimate can be calculated at the organ scale rather than at the plant scale (Chelle 2005).

Resource partitioning between plants is more sophisticated in IBMs than in crop models, with the degree of complexity often linked to the spatial resolution of the canopy and root system. For instance, light partitioning can be simulated i) at the voxel (3D pixel) scale in voxel-based IBMs that then use the Beer-Lambert law at a very fine scale of a few centimeters to meet the law's assumptions of homogeneous cover (Sinoquet et al. 2001; Munier-Jolain et al. 2013) or ii) by connecting the plant model to a surface-based radiative transfer submodel to predict light balance (transmission and absorption) inside the canopy at the organ scale (Chelle and Andrieu 1999; Röhrig et al. 1999; Barillot et al. 2014b).

As crop models, most IBMs represent belowground interactions inadequately. Except for belowground IBMs, no belowground interactions, even indirect, are included in current IBMs that simulate annual crop mixtures. Thus, the belowground IBMs applied to crop mixtures are apparently limited to a few of the FSPMs reviewed by Dunbabin et al. (2013). Most of them focus on root architecture and its impact on resource partitioning (e.g. SimRoot, Postma and Lynch 2012), leaving aside for the moment what relates to soil biota such as mycorrhizae and bacteria. More recently, the model Min3P-ArchiSimple (Gérard et al. 2017), which combines the architectural root IBM ArchiSimple (Pagès et al. 2014) and the soil reactive-transport model Min3P (Gérard et al. 2008), was used to simulate pH changes in the root zone of two representative intercropped legumes and cereals (Gérard et al. 2018).

More particularly, aboveground FSPMs ignore belowground interactions (and vice versa for belowground FSPMs) whereas the link between above- and belowground processes underlies many benefits of crop mixtures; however, predicting mixture performance based on both above- and belowground resource partitioning is an objective of the current European Union project ReMIX (https://www.remix-intercrops.eu/).

Few IBMs, however, include some crop model modules, thus connecting a detailed representation of aboveground plant parts to simpler representations of roots and soil, which can be useful when competition for light is strong or spatially heterogeneous and competition for soil resources is restricted to mobile elements (i.e. water, mineral N) or negligible compared to competition for light. For instance, soil modules of the crop model STICS are used in FLORSYS to predict soil N and water which are then uptake by individual plants, with a feedback enabling to reinitialize the soil characteristics for the day after. The opposite is observed for the model SPACSYS (Wu et al. 2007; Zhang et al. 2016), which represents canopy and soil functioning in the same way as crop models but uses an additional IBM sub-module to represent root architecture.

**4. Using modeling to study annual crop mixtures**

4.1. Different models for different purposes

Several recent reviews suggested that models simulating heterogeneous annual crop mixtures could be improved by representing complex and dynamic spatio-temporal interactions among plants, as well as phenotypic plasticity (Malézieux et al. 2009; Fletcher et al. 2016). This implicitly suggests that IBMs are more suitable than crop models for simulating crop mixtures. Making models more complex has a cost, however, and models of different resolutions are designed to address different issues. Consequently, several models that differ in the level of detail need to be used to study a given system (Evans et al. 2013; Li-li et al. 2015), with a gradient from empirical to more mechanistic models, depending on the underlying objective (Passioura 1996). Consequently, the strengths and weaknesses of crop models and IBMs are inherently complementary. Hereafter, the main issues of annual crop mixtures that each modeling approach can currently address are illustrated, when possible, with examples of model applications. These issues are related to annual crop mixtures agricultural management, and to the traits and the relative importance of ecological processes involved in crop mixtures performance.

4.2. Suitable management for annual crop mixtures: choosing species, genotypes, sowing date and density, and spatial arrangement

The spatio-temporal arrangement of annual crop mixtures influences crop yield directly through resource partitioning and indirectly through regulation of pests (Ratnadass et al. 2012; Boudreau 2013; Hatt et al. 2018). This issue is partially addressed by crop models and IBMs that simulate agricultural practices.

None of these crop models simulates more than two crops growing together, which is the most common situation for annual crop mixtures on farms. The model chosen must represent the spatial arrangement of the crop mixture simulated (closely mixed vs. relay-strip). In contrast, IBMs enable theoretical testing of all species and genotype combinations, regardless of the number of plant components and their spatial arrangement, under the condition that the IBMs represent the functions and specific characteristics of the mixed species or genotypes.

These models do not include rules to decide which species and genotypes to combine for a given soil and climate. This decision can be made by simulating many different agronomic situations to identify potential species/genotype combinations. This is possible only when the model has a minimum degree of genericity, which is the case for certain crop models (e.g. STICS and APSIM parameterize 15 and 24 annual cash crops, respectively, with genotype-specific parameterization for phenology) and IBMs (e.g. FLORSYS parameterizes 14 annual cash crops, with genotype-specific parameterization for wheat, field bean and pea). However, among IBMs, most FSPMs are species-specific.

In addition to sowing dates and densities, which most models usually represent, a specific strength of most crop models is that they consider other agricultural practices, such as fertilization, tillage, and irrigation; IBMs typically do not consider these practices. Thus, crop models have been used mainly to test the influence of a variety of agricultural practices and cropping conditions on the relative performance (yield quantity and quality) and resource-use efficiency of annual crop mixtures compared to their corresponding sole crops. For instance, simulations were performed to quantify impacts of crop mixture sowing densities and dates (Corre-Hellou et al. 2009; Fayaud et al. 2014) and to test fertilization options (Corre-Hellou et al. 2007, 2009) and water regimes (Chimonyo et al. 2016). Few crop models have been used to simulate annual crop mixtures in more integrated farming systems that have planning horizons longer than one year. Launay et al. (2009) used STICS to assess bi-specific pea-barley intercrop sown in alternate rows to improve N use efficiency by using sowing dates and densities and crop rotation (preceding crop and soil). Crop models that include a soil component and N residue management can simulate carbon and N cycles and thus indirect N transfer through decomposition and mineralization of plant residues, which could have a positive long-term effect, e.g. Fletcher et al. (2015) demonstrated that N fixation in a wheat-pea intercrop can lead to yield improvements in subsequent wheat crops.

For the additional border-row effect in relay-strip crop mixtures, simulations were performed to quantify the effect of strip width (defined by the number of rows and inter-row distance) on light use efficiency and yield of wheat-maize intercrops (Wang et al. 2015; Gou et al. 2017b). The effect of diverse crop phenology on crop yield in wheat-soybean intercrops was also evaluated under a variety of soil and climate conditions (Monzon et al. 2007).

Even though simulation studies illustrate the relevance and ability of crop models to simulate annual crop mixtures, these models are often not evaluated with independent data (e.g. Corre-Hellou et al. 2007, 2009) and thus might be over-fitted to calibration data (e.g. Chimonyo et al. 2016), which is a common problem in modeling (Sinclair and Seligman 2000; Seidel et al. 2018).

Crop models usually do not represent multiple biotic stresses, which is a clear limitation to using them to simulate low-input or organic cropping systems. One way to represent plant-pest interactions is to combine crop models with pest models, as is done for sole crops (Boote et al. 1983; Robert et al. 2004). Generally, IBMs consider biotic stresses. For example, an explicit 3D plant canopy simulated by an FSPM can help quantify the influence of the mixture on pest dynamics in homogeneous genotype mixture (e.g. Gigot et al. 2014 for wheat genotype mixture); Colbach et al. (2014b) used FLORSYS to test several crop mixtures as a function of species choice, sowing date and arrangement, focusing on weed-induced yield loss and weed biomass during the test year and the following 10 years; Potting et al. (2005) developed a simple agent-based IBM to study effects of the proportion of a sensitive crop vs. a non-sensitive crop and the spatial arrangement of the resulting mixture to predict the behavior and dynamics of herbivorous insects.

## 4.3. Plant traits involved in annual crop mixture performance

Breeding programs focus on optimizing agronomic performance (e.g. yield quantity and quality, pest resistance, resource use efficiency) in a sole crop context; however, the performance-related traits of mixtures might differ due to specific plant-plant interactions and complementarities (Litrico and Violle 2015). IBMs that are FSPMs can often identify these traits.

Several descriptive FSPMs have been developed with this objective, i.e. to replicate the plant structure (shoots or roots) observed from specific experimental measurements. Due to their detailed structural representation, these models provide insights into physical interactions between plant structure and usually a single environmental factor such as light, water, N, or pests. For instance, to simulate closely-spaced wheat-pea mixtures and determine the influence of several architectural traits on light partitioning, Barillot et al. (2014b) used a descriptive FSPM composed of a wheat model (ADEL-Wheat, Fournier et al. 2003), a pea model (L-Pea, Barillot et al. 2014b), and a radiative balance model (CARIBU, Chelle and Andrieu 1998). More specifically, their study quantified the relative influence of leaf area index (influenced by the number of branches/tillers), plant height (reflected by internode length), and leaf geometry (reflected by leaf angle) on the mixture's light interception. Similarly, but for relay-strip wheat-maize intercrops, Zhu et al. (2015) performed simulations to identify the relative importance of phenotypic plasticity in key architectural traits (e.g. tillering, leaf size, leaf azimuth, leaf angle) vs. crop structure in light interception.

For the root system, the model SimRoot was used to simulate closely-spaced dense mixtures of maize-bean-squash (Postma and Lynch 2012) to analyze the influence of architectural root traits and complementarity on crop mixture functioning and to focus on the importance of biological rhizosphere processes as a function of nutrient limitation (N, P, K).

These findings could help breeders prioritize traits and determine the data most essential to collect in field experiments. Complex traits such as height that are used in more integrated models are separated into finer traits in FSPMs, which helps to understand more mechanistically which traits are involved in resource partitioning and thus in plant performance.

Due to the descriptive nature of these models, emergent effects (such as plasticity) of crop mixtures on crop performance cannot be simulated because plant growth and development are inferred from experiments (e.g. no physiological processes are behind architectural plasticity as plasticity is hard-coded in the model), meaning that a new set of parameters is required for each situation.

## 4.4. Representation and relative importance of ecological processes

Ecological processes, such as phenotypic plasticity, facilitation, and competition are included in models that use descriptive functions (usually crop models) or dose-response type relationships (usually IBMs). In the latter, ecological processes are emergent properties of the model, i.e. the result of modeling plants' resource acquisition and signaling. Depending on how ecological processes are represented (descriptive vs. emergent), parameterization conditions greatly influence model accuracy and genericity.

Since crop models tend to use empirical relationships to represent ecological processes, using model components such as plant-related parameters developed for a sole crop might have limitations (e.g. Baumann et al. 2002; Corre-Hellou et al. 2009), and their values may change in crop mixtures due to plant-plant interactions. Consequently, parameters of model components should be estimated or measured for mixtures. Although IBMs are also parameterized from isolated plants or sole crops (except by Zhu et al. 2015), the parameterization conditions are less important since these ecological processes are supposed to be emergent properties that do not depend on the presence or identity of neighboring plants. Nevertheless, since all models are simplifications, some formalisms remain empirical, and the extent to which emergent properties are robust is often unclear, especially when simulating a variety of agricultural conditions. Thus, in addition to evaluating model accuracy for main output variables, model evaluation should also focus on ecological processes themselves. For instance, for sole crops, Casadebaig et al. (2011) used standard accuracy metrics to evaluate a crop model and showed that while the model did simulate phenotypic plasticity, its intensity was weaker than that observed in the experimental dataset.

Crop models cannot mimic phenotypic plasticity of each species (or genotype) within the mixture, but some are able to simulate cover plasticity as an emergent property of crop mixtures due to light, water, or nutrient interactions and their effects on crop growth. However, improved knowledge of the underlying processes is required, as are specific experiments to identify these mechanisms (Barot et al. 2017). This could provide lower and upper bounds of plastic functional traits that correspond to crop model parameters. Assuming that the relevant functional traits (relevant according to the issue addressed by the model) are known, that their range is quantified (phenotypic plasticity), and that they correspond to crop model parameters, the way each species/genotype expresses phenotypic plasticity could be determined by optimizing the algorithms applied to the model in the chosen environments.

More mechanistic IBMs can represent "generic" responses of plants to their environment (e.g. photosynthesis to light, or leaf growth to temperature; Barillot et al. 2016 illustrate this for wheat). In these IBMs, plant plasticity in response to contrasting environments is an emergent property. Current IBMs do not represent the functions responsible for belowground facilitation for abiotic resources, as previously defined. Some FSPMs are promising tools to simulate belowground facilitation mechanistically. For instance, the model Min3P-ArchiSimple was used to simulate adjacent legume and cereal root systems with varying distances between plants (35 and 5 cm). Preliminary results highlighted that dissolved P increased and pH decreased in the cereal's rhizosphere, while the opposite trend occurred in the legume's rhizosphere (Figure 6), depending on the distance between the two species (Gérard et al. 2018). Similarly, facilitation for P or other rhizosphere processes could be quantified based on soil conditions. Currently, only theoretical applications have been performed.

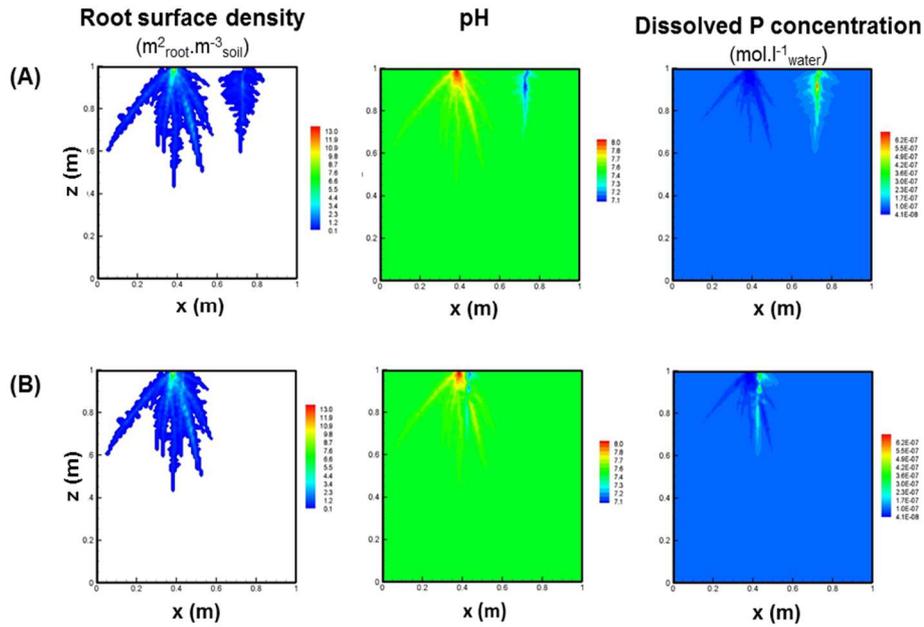

Figure 6. Results of the Min3P-ArchiSimple model used to simulate two species (cereal and legume) planted in an alkaline soil (Gérard et al. 2017). The example represents 60 days of simulation and a planting distance of (A) 35 cm and (B) 5 cm.

Mechanistic IBMs are useful to identify the relative importance of ecological processes in the functioning and performance of crop mixtures. For instance, Zhu et al. (2015) simulated a relay-strip crop mixture of wheat and maize with or without plasticity, based on empirical values of traits. The results indicated that light interception efficiency was 23% higher in crop mixtures than in sole crops, and within the 23%, 64% was due to plasticity, and only 36% resulted from crop structure. Thus, a model that does not explicitly simulate plasticity would fail to capture the main benefits of crop mixtures. Similarly, Postma and Lynch (2012) illustrated that the benefits of closely-spaced mixtures for N uptake and biomass production emerged only when root architecture was represented because they depend on spatial complementarity among roots. Likewise, modeling could help to disentangle and better analyze and understand the relative importance of the complementarity and "selection effects" in overyielding of annual crop mixtures, which remains a research perspective not yet addressed.

4.5. Complementarity of the modeling approaches

The two modeling approaches are complementary, and having suitable modeling infrastructure can increase their use. As an example, IBMs could be used to assert certain empirical assumptions used in crop models. Barillot et al. (2011) illustrated a reassuring example in a study that assessed the "homogeneous canopy" assumption for mixtures. The authors used simulation to show that the "turbid medium" analogy could be used successfully for a wide range of crop mixtures. More detailed representation (more vertical layers or 3D description) of the canopy, however, slightly improved the prediction of light interception efficiency of mixtures with vertical stratification and overlapping foliage. This study validates the assumption of the "big leaf" approach used in all crop models.

High expectations of IBMs are based on their ability to quantify the contribution of ecological processes or the importance of given functional traits to crop mixture functioning, which could encourage their use in crop models. When needed, important processes highlighted from IBMs could be summarized by functional relationships (Escobar-Gutiérrez et al. 2009) and be added to crop models. Therefore, IBMs could help improve crop models; however, as mentioned, to be parameterized and calibrated, they require many field data, which are time-consuming, difficult, or sometimes impossible to measure. A promising tool to render these models more usable is phenotyping of sole crops and crop mixtures (Bucksch et al. 2014). For instance, multi-genotype canopies of thousands plants have been simulated in 3D from a phenotyping platform to estimate light interception and light use efficiency (Cabrera-Bosquet et al. 2016; Chen et al. 2018). Similarly, IBM calibrations obtained from isolated plant phenotyping were also shown able to predict the outcome of competitive interactions between contrasting genotypes (Faverjon et al. 2018).

## 5. Conclusions and perspectives

We assert that models would be useful tools to understand and predict the functioning of annual crop mixtures and to help designing these agroecosystems, as a complement to field experiments, keeping in mind that experiments are crucial to improve and refine simulation models whereas models are useful to guide experiments and test all possible plant-environment combinations (Craufurd et al. 2013; Reynolds et al. 2018; Rötter et al. 2018). However, modeling of annual crop mixtures is still in its infancy, and most modeling studies focus on model evaluation (Gou et al. 2017b) or the evaluation of sole crop models extended to crop mixtures, in particular for bi-specific mixture of annual crops (Corre-Hellou et al. 2009; Knörzer et al. 2010; Munz et al. 2014). Use of modeling remains limited in part because models are not yet completely operational, indicating the need to increase understanding of the ecological processes that influence crop mixture functioning before simulating them. A large scientific effort has focused on modeling sole crops. For instance, many modelers have focused on defining the most relevant parameters for sole crops (Martre et al. 2015; Casadebaig et al. 2016b), while little insight exists on which parameters are particularly important for crop mixtures. A lot of work is now necessary in order to be able to identify the most relevant parameters for crop mixtures, including those linked to trait plasticity.

The complementarity between crop models and IBMs promotes stronger connection between communities of modelers and is a promising way to promote the use of annual crop mixtures in modern agriculture. Similarly, current simulation platforms are sufficiently developed to connect models. Some platforms, such as RECORD (Bergez et al. 2013) and APSIM (Holzworth et al. 2014), focus specifically on crop models (creating, connecting, simulating, and sharing them) and the cropping system scale.

Other platforms, such as OpenAlea (Pradal et al. 2015) and GroIMP (Hemmerling et al. 2008), are designed for IBMs. These platforms facilitate connection and coupling of models, using shared objects to represent plant topology (Godin and Caraglio 1998; Balduzzi et al. 2017) and geometry at multiple scales (Pradal et al. 2009; Balduzzi et al. 2017). Recent developments in platform interoperability enable processes and architectural models to be shared between platforms (Long et al. 2018); however, this work is ongoing for crop model and IBM platforms. Some platforms (OpenAlea, CAPSIS) include models that focus on multi-species communities, such as forests or grasslands, and could provide insights to improve models of annual crop mixtures.

Modeling also requires access to multiple datasets, which promotes sharing of public data (Reichman et al. 2011) and using homogenized data at the local scale to generate common databases of agronomic and environmental data. These needs are not completely met by ecological databases such as TRY (Kattge et al. 2011), which are developed mainly for natural ecosystems and species. These perspectives could be viewed as good practice for model use and sharing (Wilson et al. 2017).

Finally, modeling of annual crop mixtures were discussed in this review at the field scale, focusing mainly on yield and crop mixture functioning; however, agriculture is more and more required to be multifunctional as a way of agroecology (Gaba et al. 2018). For this reason, future models will have to simulate impacts of annual crop mixtures on various ecosystem services and the field scale should be included in more integrated scales such as the rotation and the agricultural landscape, including long-term effects.

**Acknowledgement**

The authors acknowledge support from European Union through the project H2020 ReMIX (Redesigning European cropping systems based on species mixtures, https://www.remix-intercrops.eu/) and from INRA Environment and Agronomy Division through the project IDEA (Intra- and interspecific diversity mixture in agriculture). We also thank Michael and Michelle Corson for their helpful comments and English revision and the two anonymous reviewers for their valuable comments.